\documentclass[letterpaper,twoside,12pt]{article}
\usepackage{amssymb}
\usepackage{amsmath}
\usepackage{latexsym}
\usepackage{verbatim}
\usepackage{graphicx}
\usepackage{epsfig}
\usepackage{stmaryrd}
\usepackage{fancyheadings}
\usepackage{pstricks,pst-node}
\usepackage{rotating,graphicx}
\usepackage[english]{babel}
\newcommand{\papertitle}{%
Does Quantum Mechanics Clash with the Equivalence Principle -- and Does it Matter?%
}
\newcommand{\runningtitle}{%
Does Quantum Mechanics Clash with the Equivalence Principle?%
}
\newcommand{\paperauthor}{%
Elias Okon and Craig Callender%
}
\begin{document}

%%%%%%%%%%%%%%%%%%%%%%%%%%%%%%%%%%%%%%%%%%%%%%%%%%%%%%%%%%%%%%
% Titlepage
%%%%%%%%%%%%%%%%%%%%%%%%%%%%%%%%%%%%%%%%%%%%%%%%%%%%%%%%%%%%%%
\begin{titlepage}
\vspace*{-1cm}
\begin{flushright}
\textsf{}
\\
%\textsf{ICN-UNAM-yy/pp}
\mbox{}
\\
\textsf{\today}
\\[3cm]
\end{flushright}
%%%%%%%%%%%%%%%%%%%%%%%%%%%%%%%%%%%%%%%%%%%%%%%%%%%%%%%%%%%%%%
%%%%%%%%%%%%%%%%%%%%%%%%%%%%%%%%%%%%%%%%%%%%%%%%%%%%%%%%%%%%%%
%%% TITLE, AUTHORS
%%%%%%%%%%%%%%%%%%%%%%%%%%%%%%%%%%%%%%%%%%%%%%%%%%%%%%%%%%%%%%
%%%%%%%%%%%%%%%%%%%%%%%%%%%%%%%%%%%%%%%%%%%%%%%%%%%%%%%%%%%%%%
%\begin{center}
%%%%%%%%%%%%%%%%%%%%%%%%%%%%%%%%%%%%%%%%%%%%%%%%%%%%%%%%%%%%%%
\renewcommand{\thefootnote}{\fnsymbol{footnote}}
\begin{Large}
\bfseries{\sffamily \papertitle}
\end{Large}

\noindent \rule{\textwidth}{.6mm}

\vspace*{1.6cm}

\noindent \begin{large}%%
\textsf{\bfseries%
\paperauthor
}
\end{large}

\vspace*{.1cm}

\phantom{XX}
\begin{minipage}{.8\textwidth}
\begin{it}
\noindent Philosophy Department,\\
UC San Diego,\\
9500 Gilman Dr., La Jolla, CA 92093\\[.5mm]
\end{it}
\texttt{eokon@ucsd.edu, ccallender@ucsd.edu}
\phantom{X}
\end{minipage}
\\

\vspace*{3cm}
%%%%%%%%%%%%%%%%%%%%%%%%%%%%%%%%%%%%%%%%%%%%%%%%%%%%%%%%%%%%%%
%%% ABSTRACT
%%%%%%%%%%%%%%%%%%%%%%%%%%%%%%%%%%%%%%%%%%%%%%%%%%%%%%%%%%%%%%
\noindent
\textsc{\large Abstract: }
With an eye on developing a quantum theory of gravity, many physicists have recently searched for quantum challenges to the equivalence principle of general relativity.  However, as historians and philosophers of science are well aware, the principle of equivalence is not so clear.  When clarified, we think quantum tests of the equivalence principle won't yield much.  The problem is that the clash/not-clash is either already evident or guaranteed not to exist.  Nonetheless, this work does help teach us what it means for a theory to be geometric.
\end{titlepage}
\setcounter{footnote}{0}
\renewcommand{\thefootnote}{\arabic{footnote}}
\setcounter{page}{1}
%%%%%%%%%%%%%%%%%%%%%%%%%%%%%%%%%%%%%%%%%%%%%%%%%%%%%%%%%%%%%%
%%%%%%%%%%%%%%%%%%%%%%%%%%%%%%%%%%%%%%%%%%%%%%%%%%%%%%%%%%%%%%
%%%%%%%%%%%%%%%%%%%%%%%%%%%%%%%%%%%%%%%%%%%%%%%%%%%%%%%%%%%%%%
\noindent \rule{\textwidth}{.5mm}

\tableofcontents

\noindent \rule{\textwidth}{.5mm}

%\begin{document}

%\begin{center}
%\Large{\textbf{Does Quantum Mechanics Clash with the Equivalence Principle -- and Does it Matter?}}
%\end{center}

%\vspace{0cm}

%\begin{quote}
%With an eye on developing a quantum theory of gravity, many physicists have recently searched for quantum challenges to the equivalence principle of general relativity.  However, as historians and philosophers of science are well aware, the principle of equivalence is not so clear.  When clarified, we think quantum tests of the equivalence principle won't yield much.  The problem is that the clash/not-clash is either already evident or guaranteed not to exist.  Nonetheless, this work does help teach us what it means for a theory to be geometric.
%\end{quote}

%%%%%%%%%%%%%%%%%%%%%%%%%%%%%%%%%%%%%%%%%%%%%%%%%%%%%%%%%%%%%
\section{Introduction}
%%%%%%%%%%%%%%%%%%%%%%%%%%%%%%%%%%%%%%%%%%%%%%%%%%%%%%%%%%%%%

Essential to the search for a theory of quantum gravity is a good understanding of exactly where quantum mechanics and general relativity conflict.  Rather than focus on the peripheral claims of each theory, it's natural to instead concentrate on conflicts that may arise among core principles. Hence some research has examined tensions arising from applying the principle of superposition –- surely a core aspect of quantum mechanics, if any –- to the gravitational context. However, since the principle of superposition generates notoriously perplexing conceptual difficulties, such as the quantum measurement problem, this path is fraught with complications.  More promising, perhaps, would be working in the other direction: instead of finding gravitational phenomena that conflict with a core principle of quantum mechanics, discover quantum phenomena that conflict with a core principle of relativity. Given the increasingly sensitive accuracy of quantum measurements and the comparative difficulty of gravitational measurements, this path may better yield to practical investigation too.

Motivated in part by the above reasons, there has recently emerged a literature in physics that seeks quantum challenges to the equivalence principle.1	The equivalence principle is a natural one to associate with the core of general relativity. On its face, it is more substantive than the symmetry of general covariance; that is, there are generally covariant spacetimes that do not obey the principle of equivalence.  Furthermore, general covariance finds itself almost as controversial as superposition in quantum theory. So if looking for a quantum clash with a core principle of general relativity, the equivalence principle has a lot to recommend itself.

However, as historians and philosophers of science are well aware, the principle of equivalence is itself hardly the best understood principle we have. Outside the realm of textbooks, there are almost as many equivalence principles as there are authors writing on the topic. The question naturally arises: if quantum mechanics conflicts with the equivalence principle, which one does it conflict with? Right now some papers find conflicts where others do not.  Here we hope to resolve these ambiguities by showing that different equivalence principles are at stake. By clarifying various equivalence principles and challenges, we hope to organize the messy literature on this topic. That said, our main goal is not merely clerical. The point of this literature is to find meaningful clashes between the quantum and classical. As we clarify, we evaluate the particular claims about each clash/not clash for each equivalence principle. Although there is much of interest here with respect to general relativity and the equivalence principle, an outcome of this analysis is that we'll end up very skeptical about this avenue of study. Read one way, quantum mechanics does clash with the equivalence principle, but this principle isn't in fact part of general relativity. Read as something plausibly part of general relativity, by contrast, the principle simply cannot conflict with quantum mechanics. Either way, a quantum conflict with the equivalence principle doesn't shed as much light on quantum gravity as we would hope. Nonetheless, it does, we think, teach us something about what it means to say that gravity is geometrical.

The literature contains a vast array of logically and physically inequivalent statements of the equivalence principle. We believe these principles can be neatly classified into four categories that represent the different core ideas at stake. The four groups correspond to propositions about the i) universality of free fall, ii) equivalence between homogeneous gravitational fields and uniform accelerated motion, iii) universality of behavior of physical systems in "small" regions of spacetime and iv) universal and minimal character of matter-gravitational couplings. There are of course various interconnections among the different classes, but a clear recognition of their differences is essential for an assessment of possible clashes with the quantum formalism. Let's begin with the universality of free fall.
 
%%%%%%%%%%%%%%%%%%%%%%%%%%%%%%%%%%%%%%%%%%%%%%%%%%%%%%%%%%%%%
\section{Universality of Free Fall}
%%%%%%%%%%%%%%%%%%%%%%%%%%%%%%%%%%%%%%%%%%%%%%%%%%%%%%%%%%%%%

The universality of free fall, often referred to as the \emph{weak} equivalence principle, is the claim associated with Galileo's famous experiment of dropping bodies of different mass from a great height.  It asserts that
\begin{quote}
\textbf{(EP1)} All test bodies fall in a gravitational field with the same acceleration regardless of their mass or internal composition.
\end{quote}
EP1 is exactly true in Newtonian mechanics, as it is equivalent to the requirement of equality between inertial and gravitational masses.  This principle even remains true if we drop the test character demand since the trajectory of the center of mass of any mass distribution in free fall is independent of its composition and internal structure.

In quantum mechanics, by contrast, the situation is more complicated and perhaps surprising.  The first indication of trouble comes from the demand of a test object, for this requirement poses serious problems in the quantum domain. This is because in a quantum mechanical context i) we may not be able to make the energy of a particle as small as we want in order to avoid back-reaction, ii) we cannot make the momentum of a particle as small as desired and continue to demand localization, and iii) objects may be affected by measurements and failure to recognize this may result in internal contradictions.   In addition, it is hardly clear what the ``free'' in ``free fall'' means in a quantum world. What does it mean for an extended object, a probability distribution, to be freely falling?\footnote{This problem even survives transition into quantum ontologies that do have determinate objects, such as Bohmian mechanics.~\cite{Son:95} puts Bohm's particle interpretation theory into second-order formulation to see what ``free'' Bohm particles look like, but~\cite{Val:97} rightly points out that the theory is really a first-order one.  ``Free'' Bohm particles, in this formulation, just sit still.  Rather than draw the conclusion Valentini does, namely, that Bohmian mechanics demands a spacetime with Aristotelian structure, we take this as a reductio of the approach that demands that the Newtonian framework of free versus forced motion is strictly applicable in a quantum world.}

Worse, even if these complications could be ignored, the fact remains that quantum objects do not even satisfy the essence of EP1.  That is, the behavior of quantum particles in external gravitational fields is mass dependent.  While this has been clearly recognized in particular situations, e.g., the COW experiments~\cite{Col.Ove.Wer:75}, it often has been considered an atypical behavior.  In any case, this mass-dependent behavior is not that uncommon and should come as no surprise taking into account that the behavior of even \emph{free} quantum particles is mass dependent.  To see this consider the fact that, contrary to what happens in Newton's law for a free particle, ma = 0, the mass does not drop from the corresponding Schr\"odinger equation
\begin{equation}
i \hbar \, \frac{\partial \psi}{\partial t} = -\frac{\hbar^2}{2 m} \nabla^2 \psi \, .
\end{equation}
For the specific case of quantum particles in uniform gravitational fields, by invoking Ehrenfest's theorem it is clear that the mean time taken by the particle to fall is equal to what would be expected classically. Nevertheless, an estimate of fluctuations around this mean value can also be calculated~\cite{Vio.Ono:04,Ali.Maj.Hom.Pan:06}, yielding a standard deviation proportional to $\hbar/m$, thus signaling the non-universality of quantum free fall. Therefore, the time of descent of very light particles is expected to deviate greatly from the Newtonian value but the behavior of heavy objects is, reassuringly, expected to approach the classical, mass-independent result.

What are we to make of the fact that quantum mechanics does not satisfy the universality of free fall?  Of course, if the equivalence principle is understood as EP1 then quantum mechanics violates it.  But is this interesting or relevant to the search for quantum gravity? One point of view would be to say that it is because it undermines the motivation for thinking of gravity as geometric (see for example~\cite{Ana:80,Son:95,Ahl:97,Ald.Per.Vu:05}). The fact that all classical bodies move in the same way in a gravitational field, says the conventional wisdom, is what allows for a geometric description of the gravitational field. Nevertheless, although this type of reasoning is natural, we do not think it is correct. First, at most it would remove one reason for thinking it geometric, not provide a reason for thinking it not. Second, it does not even accomplish that, for notice that EP1 does not even hold exactly in general relativity, the archetypal geometrical theory. The point is that the geodesic principle, i.e., the fact that free objects follow spacetime geodesics, is not valid in general, but can only be exactly proven for special systems like very light (i.e., non-back-reacting) point-particles~\cite{Ger.Jan:75}, and not even in that case if certain energy conditions are not satisfied~\cite{Mal:09}. And since classical point-particles can have no internal structure whatsoever (except mass), in general relativity EP1 is not satisfied independently of internal structure but only for particles with no such thing as internal structure. Quantum mechanical point-particles on the other hand can of course possess spin, and as is well known spinning point-objects do not follow geodesics~\cite{Pap:51}. As for extended objects, it is not clear what it means for them to satisfy the geodesic principle. One way to interpret it would be to check if the center of mass of the distribution follows a geodesic but note that the definition of center of mass in general relativity is notoriously difficult and controversial~\cite{Dix:64,Bei:67,Mad:69,Ada.Koz.New:09}.

It is interesting to note that both in quantum mechanics and general relativity the universality of free fall is not true in general but is recovered in certain limits. However, the limits in which it is recovered in each case are in a sense opposite since in general relativity we need to consider light particles to avoid back-reaction, but in quantum mechanics those are precisely the particles that deviate the most from an exact massless behavior.

Finally, another reason to mistrust the link between universality of free fall and geometricity of spacetime is given by the fact that in some cases the universality of free fall is not sufficient for geometrization. The universality of free fall guaranties that the path of free particles is independent of internal structure; however, it's also important that other forms of dependence, for example velocity, be avoided too. To illustrate this point, consider a region of spacetime with a given background electromagnetic field, and suppose that in order to probe it we only have at our disposal test particles of fixed $q/m$ ratio. It follows from the Lorentz force law that the trajectories of all such particles will satisfy a toy version of the universality of free fall: trajectories will be equal for any two particles sharing initial position and velocity, independently of structure and composition (of course, except for the part that fixes the $q/m$ ratio). The question is then if it is possible to describe the trajectories of the available test particles as geodesics of a certain curved spacetime, as is done with gravity? Is it the case that the restricted universality of free fall implies that in this scenario the electromagnetic field can be geometrized? The answer is no. This is because, according to the Lorentz force law, the acceleration suffered by a particle depends on its velocity. Therefore, as opposed to the gravitational case, no family of inertial frames can be associated with each spacetime point. That is, there is no state of motion with respect to which all of our test particles at a given point are not accelerating, which is exactly what happens with gravity.  Universality of free fall does not imply the ability to geometrize a force.\footnote{Notice that the above argument could seem to contradict the existence of gravitomagnetic forces, i.e., the fact that in linearized general relativity and when the lowest order effects of the motion of the sources are considered, the geodesic equation for test bodies closely resembles the Lorentz force law (see~\cite{Wal:84}). How could the  Lorentz force law prevent geometrization if it is present in general relativity itself? Note however that whereas in the electromagnetic case the appearance of the Lorentz force law is taken to be \emph{fundamental} its appearance in general relativity is only an artifact of the approximation, in particular of setting $v^0 = dx^0/d\tau \approx 1$.}

Yet another motive for not counting the lack of universality of quantum free fall as an argument against the geometric picture of gravity is to note that the mass-dependence of freely falling quantum particles can be understood on the basis of kinematical arguments. This takes us to the second class of equivalence principles above, those referring to a full equivalence between homogeneous gravitational fields and uniform accelerated motion.

%%%%%%%%%%%%%%%%%%%%%%%%%%%%%%%%%%%%%%%%%%%%%%%%%%%%%%%%%%%%%
\section{Einstein's Equivalence Principle}
%%%%%%%%%%%%%%%%%%%%%%%%%%%%%%%%%%%%%%%%%%%%%%%%%%%%%%%%%%%%%

In Newtonian mechanics, the mass-independence described by EP1 implies that in the presence of a uniform gravitational field Newton's equations can be dramatically simplified. In fact, it allows us to introduce coordinates that remove the effects of gravity completely.\footnote{Newton's second law and law of gravitation tell us that $m_j \ddot{\mathbf{x}}_j = \sum_i \mathbf{F}_{ij} + M_j \mathbf{g}$ , where $m$ is the inertial mass, $M$ the gravitational mass, $\mathbf{g}$ a constant gravitational acceleration, and $\mathbf{F}_{ij}$ the net force of particle $i$ on $j$.  Due to the weak equivalence principle, $m=M$, and letting $\mathbf{x}'_j = \mathbf{x}_j - \frac{1}{2} \mathbf{g} t^2$ , one eliminates the gravitational field: $m_j \ddot{\mathbf{x}}_j = \sum_i \mathbf{F}_{ij}$ . }   Wanting something similar (see~\cite{Nor:85}), Einstein formulated: 
\begin{quote}
\textbf{(EP2)} A state of rest in a homogeneous gravitational field is physically indistinguishable from a state of uniform acceleration in a gravity-free space.
\end{quote}
Note that, unlike EP1, this statement constrains the totality of physical systems, not only the mechanical behavior of very special systems, i.e., free test particles. As such, it is a more general principle and potentially of greater relevance to an assessment of the geometric character of gravity at the quantum level.

From the point of view of EP2, the universality of free fall, i.e., the fact that the behavior of \emph{classical} particles in a gravitational field is mass-independent, must be considered a logical consequence of the important but seemingly innocuous fact that the behavior of \emph{free} classical particles is mass-independent -- note that there is no need to introduce mass in order to enunciate Newton's first law. The reasoning goes like this. The behavior of free particles is mass-independent and remains so when observed from an accelerated reference frame. Now, if Einstein's principle obtains, the same mass-independent behavior must be expected in a homogeneous gravitational field. On the other hand, if the behavior of free quantum particles is mass-dependent then it is only natural -- again, if Einstein's principle holds -- to expect a quantum departure from universal free fall. From this perspective the universality of free fall ceases to be a fundamental aspect of the equivalence principle. It is instead recognized as a mere instance of EP2 applied to a specific law -- namely, the mass-independent behavior of classical particles. EP2 stipulates an equivalence of behavior for \emph{all} physical laws and, in particular, it constrains the behavior of freely falling classical particles. Entities satisfying different laws, quantum particles for example, may not at the same time satisfy universality of free fall and Einstein's equivalence principle. 

In this light it seems misleading to call EP1 the weak equivalence principle. That name suggests that it must be logically contained in other versions. Nevertheless, as we have shown, EP2 is independent of the so-called weak equivalence principle. That is, it could be valid even if there are no systems in the world satisfying universality of free fall.

Is EP2 satisfied n quantum mechanics? In the last section we saw that quantum free fall is mass-dependent. For EP2 to hold the mass dependence must be such that it is the same as that observed by an accelerated observer for a free quantum particle. Interestingly, this is precisely the case! This assertion can be demonstrated in general by the fact that Schr\"odinger's equation for a particle in a homogeneous gravitational field of strength a,
\begin{equation}
i \hbar \, \frac{\partial \psi}{\partial t} = -\frac{\hbar^2}{2 m} \nabla^2 \psi + m  a  z \, \psi \, ,
\end{equation}
gets transformed, via the coordinate transformation to an accelerated frame
\begin{eqnarray}
z & = & z'+ v t' +\frac{1}{2} a t'^2   \\
t & = & t' \, , 
\end{eqnarray}
to a free particle equation
\begin{equation}
i \hbar \, \frac{\partial \psi'}{\partial t'} = -\frac{\hbar^2}{2 m} \nabla'^2 \psi' \, ,
\end{equation}
with\footnote{The phase factor in this transformation is mass-dependent and this fact might be viewed with suspicion.  The mass-dependence is clearly a good thing for EP2, as John Earman reminds us, since it leads to a superselection rule for mass in non-relativistic quantum mechanics.  That is, a coherent superposition of states of different mass is prohibited and this prohibition protects EP2 from possible violations due to interference effects between particles of different mass.  On these points see~\cite{Gre:79}.  Still, one might wonder what warrants the particular form of this transformation?  One cannot justify it as one does the corresponding phase factor when transforming from one inertial reference frame to another, i.e., by demanding Galilean invariance.  We acknowledge this deficiency and can point only to the fact that the particular transformation works (is compatible with the experimental evidence).} $\psi' = e^{i \alpha(z,t)} \psi$ (see~\cite{Gre.Ove:79}). Of course the non-trivial result is that the gravitational interaction can be treated in the non-relativistic quantum context as just another external potential, i.e., by the introduction in Schr\"odinger's equation of the term\footnote{For similar points of view see~\cite{Bor.Sch:79,Sta.Wer.Col.Ove:80,Chr.Sud:02,Unn:02,Her.Waw:03,Hue.Sah.Soc:06}.} $m  a  z \, \psi$. This important fact has been confirmed experimentally in, e.g., COW-type settings~\cite{Bon.Wor:83}.  Then, in so far as the equivalence principle is understood in Einstein's terms, quantum mechanics does not violate it. Whereas the COW experiments were considered evidence against the equivalence principle, understood as EP1, they turn out to be favorable confirmation of the equivalence principle, understood as EP2. 

As we pointed out above, the conformity of quantum mechanics with EP2 is to be decided solely by experiment. EP2 constrains experimental results and not our laws of physics (cf. EP4 below). The point that we are trying to stress here is that i) all experiments performed so far corroborate it and ii) no clash between quantum mechanics and E2 can be found theoretically.  

We think of the ability to fully predict a system's behavior in a homogeneous gravitational field by simply considering its behavior when uniformly accelerated as the real empirical pattern behind the equivalence principle. Stepping back, this result is a bit curious. Naively, one might have thought that the ability to get the same results in these two situations was due ultimately to the universality of free fall. But that is not right, since, as we have seen, EP1 does not hold in quantum mechanics -- quantum free fall is mass-dependent. Yet surprisingly this mass-dependence does not preclude the obtaining of this empirical pattern and hence EP2.  In retrospect, given that the universality of free fall does not hold exactly in general relativity either, and yet general relativity motivates the equivalence principle, it's clear that the universality of free fall had better not have been the source of this empirical pattern.

In any case, EP2 is very limited.  The restriction to homogeneous gravitational fields means that it is not a core principle of general relativity.  For this reason EP2 is not going to be of deep relevance to quantum gravity, inasmuch as that theory is concerned with reconciling our best theory of homogeneous and inhomogeneous gravitational fields with quantum matter. Conflict with the empirical pattern of course would be interesting but this conflict isn't predicted by quantum mechanics. 

Can EP2 be extended to more general situations?  Here physicists, historians and philosophers have been frustrated. A plethora of principles have been proposed, with little agreement resulting.  In what follows we'll examine some of the paths that have been explored.
 
%%%%%%%%%%%%%%%%%%%%%%%%%%%%%%%%%%%%%%%%%%%%%%%%%%%%%%%%%%%%%
\section{Small Enough Regions}
%%%%%%%%%%%%%%%%%%%%%%%%%%%%%%%%%%%%%%%%%%%%%%%%%%%%%%%%%%%%%
 
General relativity assumes that spacetime can be represented as a pseudo-Riemannian manifold with Lorentzian signature.  On such a manifold, the tangent space to any point is Minkowski spacetime, the flat spacetime of gravity-less special relativity.  This fact, coupled with the motivation of extending EP2 to generic spacetimes, suggests the natural strategy of trying to ``homogenize'' inhomogeneous fields by going local.  As one looks at smaller patches of spacetime, inhomogeneous fields appear more homogeneous.  In the infinitesimal neighborhood of a point, an inhomogeneous field is homogeneous.  Perhaps by restricting the principle to such regions, the thought is, one can obtain a universally valid statement that is true for variably-curved spacetimes.

It's intuitive to think that as an extended body shrinks the effects of curvature decrease.~\cite{Oha:77}, however, brings out a consideration that initially points against any ``going local'' strategy. He notes that one can in principle ``feel'' the non-vanishing of the Riemann tensor even at a point.  Consider a drop of liquid without surface tension in free fall.  When in the presence of gravity, the drop's shape will not be spherical but instead display the familiar ``bulge'' of tidal distortion.  Ohanian shows that this distortion doesn't go to zero as the droplet's radius approaches zero.  

Some commentators complain that Ohanian's example, when properly conceived as a process that takes time, is actually not very local.  But this immediately raises the question of what is.  Responding to this worry suggests a strategy of making the equivalence principle relative to certain measurement scales.  

The idea behind this approach is that the physical behavior of a freely falling system can be made, to a given accuracy, universal, by making its size sufficiently small.  Hence:
\begin{quote}
\textbf{(EP3)} For every experimental apparatus with some limiting accuracy, and for every spacetime event, there exists a (spacetime) neighborhood of that event such that the outcome of any measurement within that region with the experimental apparatus in free fall, is independent of the event and the velocity of the apparatus.\footnote{It is common with equivalence principles of this type to distinguish between cases including or excluding gravitational experiments. All we say about EP3 here holds either way.}
\end{quote}
Arguably, EP3 is both interesting and true. In a small enough region of spacetime and with some limiting accuracy, we won't be able to detect the tidal distortions of Ohanian's drop.  Or put differently, if we did detect tidal distortions, then the system under consideration, the drop, is too large to fit the authorized region, in which case EP3 doesn't have to hold. Our epistemic capacities help ``homogenize'' the inhomogeneous fields, allowing us to extend EP2. Note however that this epistemic element does not leave EP3 empty since it entails the highly non-trivial claim that, given a limiting accuracy and a local spacetime curvature scale, we can \emph{always} find a region where it holds, i.e., where the physical behavior of a freely falling system can be made universal.

Despite its non-triviality, we don't think EP3 is capable of providing a meaningful test of the conflict between general relativity and the quantum world. This is because the accuracy and size limitations that insulate EP3 from Ohanian's drop also insulate EP3 from any quantum challenge. The point is that for any quantum experiment that confronts the equivalence principle there are two options: either the quantum system fits inside the region authorized by EP3, in which case we are back to a situation analogous to the one encountered by EP2 (a clash could be found experimentally but it will not arise theoretically) or the system does not fit the region, in which case EP3 simply becomes mute. This isn't to say that this last case couldn't be useful. Experiments that attempt to detect gravitational tidal effects using quantum mechanical probes have been proposed lately~\cite{Chi:03,Sud:05},  holding the promise to bring clues about the character of gravity at the quantum level. 

What goes for Ohanian's drops goes for quantum systems.  The ``small enough'' reading of the equivalence principle, with the stress on ``enough,'' guarantees compatibility between quantum systems and the equivalence principle.  Read as EP3, quantum mechanics can't pose a challenge to the equivalence principle.

%%%%%%%%%%%%%%%%%%%%%%%%%%%%%%%%%%%%%%%%%%%%%%%%%%%%%%%%%%%%%
\section{Couplings}
%%%%%%%%%%%%%%%%%%%%%%%%%%%%%%%%%%%%%%%%%%%%%%%%%%%%%%%%%%%%%

Another reaction to the considerations brought out at the beginning of the last section is to formalize the notion of ``special relativity holding'' at a single spacetime point.  If one literally means special relativity holds then this isn't defensible (since that would include the vanishing of the Riemann tensor), but perhaps a different sense can be specified in which it is.  Consider, for instance,
\begin{quote}
\textbf{(EP4)} All nongravitational fields couple to a single gravitational field (universality), and at each point of spacetime it is possible to find a coordinate transformation such that the gravitational field variables can be eliminated from the field equations of matter (minimal coupling).
\end{quote}
Like EP3, we believe that a good argument can be made for EP4's truth and interest.  EP4 is also a substantive principle.  There are two claims here, universality and minimal coupling.  Universality insists that all matter fields couple to a single gravitational field.  This demand rules out bimetric theories of gravity, for instance.  Minimal coupling is the addition that specifies what ``special relativity holding'' means.  On our formulation, this rules out logically possible laws such as 
\begin{equation}
\nabla^a \nabla_a \phi - m^2 \phi - \xi R \phi = 0
\end{equation}
where $\phi$ is a scalar field, $R$ the Ricci scalar and $\xi$ a coupling constant.  One could insist that the equations actually look like a proper subset of those holding in special relativity, as in~\cite{Ghi.Bud:01}, but we have chosen the less restrictive form that merely demands the possible elimination of the gravitational field variables from the field equations of matter.  The intuition behind EP4 is that all non-gravitational fields ``feel'' the same gravitational field, but they do so weakly, so that at each point of spacetime it is possible to find a coordinate transformation such that neither the Riemann curvature tensor nor its contractions appear in the laws.

We pause to note that universality and minimal coupling are logically independent.  One can imagine (and sometimes finds) equivalence principles referring solely to one or the other component.\footnote{~\cite{And:67} version is essentially universality;~\cite{Oha:77} is essentially minimal coupling.}   In addition, although we prefer EP4's formulation, there are many distinct ways of understanding minimal coupling.  Finally, either of these principles can be combined with the measurement accuracy clause of EP3.  A combinatorial explosion of possible principles threatens.  Nonetheless, we believe that what we say about EP1-EP4 applies, mutatis mutandis, to the other combinations envisioned.

Although EP4 describes some deep features of our theories, it is a bit disappointing as regards any possible quantum challenge.  The problem is that EP4 builds in compatibility with the quantum.\footnote{This also happens in Ghins and Budden's principle DEEP, which requires that all the ``fundamental dynamical equations'' hold.  Since the dynamical equations of motion in quantum mechanics are such equations, it's automatic that they are compatible.} Note that it refers only to the form of the laws of physics but remains silent about experimental results. The principle demands that all quantum fields feel the same gravitational field but have dynamical equations of motion that make no essential reference to $R_{abcd}$.  Well, that's true, and it's no accident.  The pseudo-Riemannian metric $g_{ab}$ is felt by all the quantum fields; indeed, there exist deep connections between the form of the dynamical equations and this single metric, e.g., between the Lorentzian signature of the metric and the hyperbolic form of the partial differential equations describing the evolution.  And of course none of the quantum fields evolve as a function of $R_{abcd}$.\footnote{Investigations of quantum non-minimal couplings could certainly be very fruitful, but they lie outside of standard quantum mechanics and thus the scope of the present paper.} The project of searching for a quantum challenge to the equivalence principle is over and done with as soon as EP4 is written!

%%%%%%%%%%%%%%%%%%%%%%%%%%%%%%%%%%%%%%%%%%%%%%%%%%%%%%%%%%%%%
\section{Conclusion}
%%%%%%%%%%%%%%%%%%%%%%%%%%%%%%%%%%%%%%%%%%%%%%%%%%%%%%%%%%%%%
 
We have shown that, despite all the natural motivation, a search for quantum challenges to the equivalence principle seems unlikely to yield much. Read as EP1, we already know of the clash, but we also know that EP1 may not be a part of general relativity, strictly conceived, nor is it really a part of traditional ``Einsteinian'' equivalence principles.  Read as Einstein's EP2, the principle is too limited to be a core principle of relativity.  As for the sharpened versions EP3 or EP4, we either know that quantum mechanics doesn't predict a clash or that no clash is guaranteed.

Still, perhaps we've learned something from this exercise. It's natural to think of the universality of free fall as expressing the sense in which gravity is geometrical. But in general relativity properly conceived, it seems conventional what to call gravity and not gravity anyway (see~\cite{Nor:85}). If there is something to the idea -- and this is yet another posited equivalence principle -- all we mean is that all the dynamical matter fields ``feel'' the same metric.  Quantum mechanics won't challenge this, but maybe quantum gravity will.

%%%%%%%%%%%%%%%%%%%%%%%%%%%%%%%%%%%%%%%%%%%%%%%%%%%%%%%%%%%%%%
%%%%%%%%%%%%%%%%%%%%%%%%%%%%%%%%%%%%%%%%%%%%%%%%%%%%%%%%%%%%%%
%%%%%%%%%%%%%%%%%%%%%%%%%%%%%%%%%%%%%%%%%%%%%%%%%%%%%%%%%%%%%%
%\bibliographystyle{apalike} %ajae
%\bibliography{biblio}
%%%%%%%%%%%%%%%%%%%%%%%%%%%%%%%%%%%%%%%%%%%%%%%%%%%%%%%%%%%%%%

\end{document}